\documentstyle[prl,aps,preprint,psfig]{revtex} 
\begin{document}\hbadness=10000
\def\NZ{{\bar N_0}}
\def\NJ{N_{\J}}
\def\J{J/\psi}
\title{Enhanced $\J$ Production in Deconfined Quark Matter
}
\author{Robert L. Thews, Martin Schroedter,  and Johann Rafelski}
\address{Department of Physics,
        University of Arizona,
        Tucson, AZ 85721, USA}
%
%
\maketitle

\begin{abstract}\noindent 
{In high energy heavy ion collisions at the Relativistic Heavy
Ion Collider (RHIC) at Brookhaven and the Large Hadron
Collider (LHC) at CERN,  
each central event will contain multiple pairs of 
heavy quarks.
If a region of deconfined quarks and gluons is 
formed, a new mechanism for the formation of heavy quarkonium
bound states will be activated.  This is a result of the
mobility
of heavy quarks in the deconfined region, such that 
bound states can be formed from a quark and an antiquark which
were originally produced in separate incoherent interactions. 
Model estimates of this effect for $J/\psi$ production at
RHIC indicate that significant enhancements 
are to be expected.  Experimental observation of such enhanced
production would provide evidence for deconfinement unlikely to
be compatible with competing scenarios. 
PACS number(s):  12.38.Mh, 25.75.-q, 14.40Gx.
}
\end{abstract}


Ultrarelativistic heavy ion collisions at the RHIC and LHC
colliders are expected to provide initial energy
density sufficient to initiate
a phase transition from normal hadronic
matter to deconfined quarks and gluons \cite{phasetransition}.
A decrease in the number of observed heavy quarkonium states was 
proposed many years ago \cite{matsuisatz} as a signature of the deconfined phase.
One invokes the argument that in a plasma of free quarks and gluons the
color forces will experience a Debye-type screening. 
Thus the quark and antiquark
in a quarkonium bound state will no longer be subject to a confining force
and diffuse away from each other during the lifetime of the quark-gluon plasma.
As the system cools and the deconfined phase disappears, these heavy
quarks will most likely form a final hadronic state with
one of the much more numerous light quarks.  The result will be a
decreased population of heavy quarkonium relative to those formed initially
in the heavy ion collision.  

There is now extensive data on charmonium production 
using nuclear
targets and beams.
The results for $J/\psi$ 
in p-A collisions and also from 
Oxygen and Sulfur beams on Uranium show a systematic nuclear dependence
of the cross section
\cite{nucalpha}
which points toward
an interpretation in terms of interactions of an initial quarkonium
state with nucleons\cite{gerschel}.  Recent results for 
a Lead beam and target
reveal an additional suppression of about 25$\%$, prompting claims that
this effect could be the expected signature of deconfinement\cite{NA50}.  The increase
of this anomalous suppression with the centrality of the collision, as
measured by the energy directed transverse to the beam, shows signs of
structure which have been interpreted as threshold behavior due to
dissociation of charmonium states in a plasma\cite{Nardisatz}.  
However, several 
alternate scenarios
have been proposed which do not involve deconfinement effects
\cite{alternate}.
These models are difficult to rule out at present, since there is significant
uncertainty in many of the parameters.  It appears that a precision 
systematic study
of suppression patterns of many states in the quarkonium systems will be
necessary for a definitive interpretation.

In all of the above, a tacit assumption has been made:  the heavy 
quarkonium is formed only during the initial nucleon-nucleon collisions.
Once formed, subsequent interactions with nucleons or final state 
interactions in a quark-gluon plasma or with other produced hadrons can
only reduce the probability that the quarkonium will survive and be
observed.  Here we explore a scenario which will be realized
at RHIC and LHC energies, where the average number of heavy quark pairs 
produced in the initial (independent and incoherent) nucleon-nucleon
collisions will be substantially above unity for a typical central heavy ion
interaction.  Then {\it if and only if a space-time region of 
deconfined quarks and gluons is present},  
quarkonium states will be formed from combinations of
heavy quarks and antiquarks which
were initially produced in different nucleon-nucleon collisions.

This new mechanism of heavy quarkonium production has the potential
to be the dominant factor in determining the heavy quarkonium population
observed after hadronization.  To be specific, let $N_0$ be the number
of heavy quark pairs initially produced in a central heavy ion
collision, and let $N_1$ be the number
of those pairs which form bound states in the normal confining
vacuum potential.  The final number $N_B$ of bound 
states surviving at hadronization  
will be some fraction $\epsilon$ of the initial number $N_1$, plus
the number formed by this new mechanism from the remaining 
$N_0 - N_1$ heavy quark pairs.  (We include in the new mechanism both
formation and dissociation in
the deconfined region.)  The instantaneous formation rate of $N_B$ 
will
be proportional to the square of the number of 
unbound quark pairs, which we approximate by its initial value.
This is valid as long as $N_B << N_0$, which
we demonstrate is valid in our model calculations.   
We also show in our model calculations that the time scales for the
formation and dissociation processes are typically somewhat larger than
the expected lifetime of the deconfined state.  Thus there will be 
insufficient time for the relative populations of bound and unbound
heavy quarks to reach an equilibrium value, and we anticipate that the
number of bound states existing at the end of the deconfinement lifetime 
will remain proportional to the square
of the initial unbound charm population. We  introduce a 
proportionality parameter $\beta$, to express the final population as 
  
\begin{equation}
N_B = \epsilon N_1 + \beta (N_0 - N_1)^2.
\label{eqquad}
\end{equation}

We then average over the distributions of $N_1$ and $N_0$, 
introducing the probability $x$ that a given heavy quark pair
was in a bound state before the deconfined phase 
was formed.  

The bound state ``suppression" factor $S_B$ is just the ratio of
this average population to the average initially-produced
bound state population per collision, $x \NZ$.

\begin{equation}
S_B = \epsilon + \beta (1-x) + \beta {(1-x)^2\over x} (\NZ +1)
\label{eqsupp}
\end{equation}

Without the new production mechanism, $\beta = 0$ and the 
suppression factor $S_B$ is bounded by $\epsilon < 1$.  
But for sufficiently large values of  
$\NZ$ this
factor could actually exceed unity, i.e. one would predict
an {\bf enhancement} in the heavy quarkonium production
rates to be the signature of deconfinement!
We thus proceed to estimate expected $\beta$-values
for $\J$ production at RHIC.

Let us emphasize at the outset that we are not attempting a 
detailed phenomenology of $\J$ production at
RHIC.  The goal is  merely to estimate if this new formation mechanism could
have a significant impact on the results.
We consider the dynamical evolution of the $c{\bar c}$ pairs which have
been produced in a central Au-Au collision at $\sqrt{s}$ = 200A GeV.
This is adapted from our previous calculation of
the formation of $B_c$ mesons.
\cite{BC}.
For simplicity, we assume the deconfined phase is an ideal gas of free
gluons and light quarks.  To describe the ``standard model" scenario for
suppression of $\J$ in the deconfined
region, we utilize the collisional dissociation via
interactions with free thermal gluons.  (This is 
the dynamic counterpart of the static plasma screening scenario
\cite{Kha}.) 
Our new formation mechanism is just the inverse of
this dissociation reaction, when a free charm quark and antiquark 
are captured in the
$\J$ bound state, emitting a color octet gluon.  
Thus it is an unavoidable consequence 
in this model of quarkonium suppression that
a corresponding mechanism for quarkonium production must be present.
The competition
between the rates of these reactions integrated over the lifetime of
the QGP then determines the final $\J$ population.  
Our estimates result from numerical solutions of the kinetic rate equation

\begin{equation}\label{eqkin}
\frac{d\NJ}{d\tau}=
  \lambda_{\mathrm{F}} N_c\, \rho_{\bar c } -
    \lambda_{\mathrm{D}} \NJ\, \rho_g\,,
\end{equation}                                                                                                                  
where $\tau$ is the proper time,
 $\rho$ denotes 
number density,
and the reactivity $\lambda$ is
the reaction rate $\langle \sigma v_{\mathrm{rel}} \rangle$
averaged over the momentum distribution of the initial
participants, i.e. $c$ and $\bar c$ for $\lambda_F$ and
$\J$ and $g$ for $\lambda_D$.
Formation of other states containing charm quarks is expected
to occur predominantly at hadronization, 
since their lower
binding energies prevents them from existing in a hot QGP, or
equivalently they are ionized on very short time scales.

The gluon density is determined by the equilibrium value in the 
QGP at each temperature.  
Initial charm quark numbers are given by $N_0$ and $N_1$, and
exact charm conservation is enforced throughout
the calculation. 
The initial volume at
$\tau = \tau_0$ is
allowed to undergo longitudinal expansion 
$V(\tau) = V_0 \tau/\tau_0$.
The expansion is taken to be isentropic, $VT^3$ = constant,
which then provides a
generic temperature-time profile. 
We use parameter values for thermalization
time $\tau_0$ = 0.5 fm, initial volume $V_0 = \pi R^2\tau_0$ with
R = 6 fm, and a range of initial temperature 300 MeV $< T_0 <$ 500 MeV,
which are all
compatible with expectations for a central collision at RHIC.
For simplicity, we assume the 
transverse spatial distributions are uniform, and use a thermal 
momentum distribution 
for gluons.  Sensitivity of the results to these parameter values and
assumptions will be presented later.

The formation rate for our new mechanism has significant sensitivity to 
the charm quark momentum distribution, and we thus consider a wide
variation for this quantity.
At one extreme, we use the initial charm quark rapidity interval and transverse
momentum spectrum unchanged from the perturbative QCD production processes.
We then allow for energy loss processes in the plasma by reducing the width
of the rapidity distribution, terminating with the opposite extreme when
the formation results are almost identical to those which would result
if the charm quarks were in full thermal equilibrium with the plasma.
This range approximately corresponds  to changing the rapidity interval
$\Delta$y between one and four units.

We utilize a cross section for the dissociation of $\J$
due to collisions with gluons
which is based on the operator product expansion
\cite{OPE}:
\begin{equation}
\sigma_D(k) = {2\pi\over 3} \left ({32\over 3}\right )^2
\left ({2\mu\over \epsilon_o}\right )^{1/2}
{1\over 4\mu^2} {(k/\epsilon_o - 1)^{3/2}\over (k/\epsilon_o)^5},
\label{eqsigma}
\end{equation}
where $k$ is the gluon momentum, $\epsilon_o$ the binding energy,
and $\mu$ the reduced mass of the quarkonium system.  This form
assumes the quarkonium system has a spatial size small compared
with the inverse of $\Lambda_{QCD}$, and its bound state
spectrum is close to that in a nonrelativistic Coulomb potential.
This same cross section is utilized with detailed balance factors
to calculate the primary formation rate for the capture of 
a charm and anticharm quark into the $\J$.  

We have also considered a scenario in which a static screening of the
color force replaces the gluon dissociation process, and dominates the
suppression of initially-produced $\J$.  Equivalently, the binding 
$\epsilon_o$ decreases from its vacuum value at low temperature and
vanishes at high temperature.  As a simple approximation to this
behavior, we multiply the vacuum value by a step function at some 
screening temperature
$T_s$, such that total screening is active at high temperature and
the formation mechanism is active at low
temperatures.  The numerical results for these two scenarios are
identical for a screening temperature $T_s$ = 280 MeV. The screening
scenario predictions fall somewhat below the gluon dissociation
results for lower $T_s$.  They differ
by a maximum factor of two when $T_s$ decreases to 180 MeV (we have
used a deconfinement temperature of 150 MeV).

\begin{figure}[t]
\centerline{  \hspace*{-.cm}
\psfig{width=8.5cm,figure=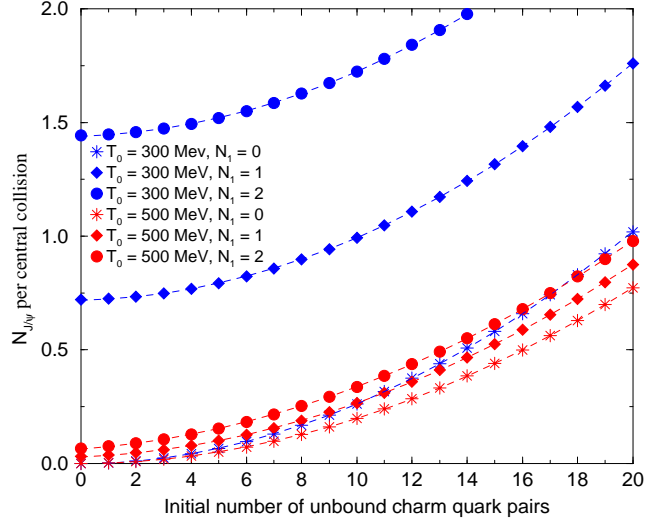}
}
\caption{ \small
Calculated $\J$ formation in deconfined matter  at several
initial temperatures, $\Delta$y = 1,
for central collisions at RHIC as a
function of initial charm pair ($N_0$)and $\J$ ($N_1$) values.
\label{figquad}}
\end{figure}

We show in Fig. \ref{figquad} sample calculated values of $\J$ 
per central event 
as a function of initial number
of unbound charm quark pairs.  
Quadratic fits of Eq. \ref{eqquad}
are superimposed. This is a direct verification of our expectations 
that the final $\J$ population
in fact retains the quadratic dependence of the initial formation rate.
This also verifies that the decrease in
initial unbound charm is a small effect. (These fits also contain a 
small linear term
for the cases in which $N_1$ is nonzero, which accounts for
the increase of the unbound charm population when dissociation occurs.)
We then extract the fitted parameters over our assumed range of initial
temperature and charm quark rapidity width.
The fitted $\epsilon$ values decrease quite rapidly with increasing
$T_0$ as expected, and are entirely insensitive to $\Delta y$.  
The corresponding
$\beta$ values have a significant
dependence on $\Delta $y. They are less sensitive to $T_0$, but
exhibit an expected decrease at large $T_0$ due to
large gluon dissociation rates at initial times (counterpart of color
screening). 

These fitted parameters 
must be supplemented by values of $x$ and $\NZ$ to 
determine the ``suppression" factor from Eq. \ref{eqsupp} for the
new mechanism.  We use the nuclear overlap function $T_{AA}$(b=0)
 = 29.3 $mb^{-1}$ for Au, and a pQCD estimate of the
charm production in p-p collisions at RHIC energy
$\sigma (pp -> c {\bar c})$ = 350 $\mu$b \cite{hardprobes1} to
estimate $\NZ$ = 10 for central collisions.

The parameter
$x$ contains the fraction of initial charm pairs which
formed $\J$ states before the onset of deconfinement. 
Fitted values from a color evaporation model 
\cite{hardprobes2} are consistent with $10^{-2}$, which we adopt 
as an order of magnitude estimate.
This must be reduced by  
the 
suppression due to interactions with target and beam nucleons.  For
central collisions we use 0.6 for this factor, which results 
from the extrapolation of the observed nuclear effects for
p-A and smaller A-B central interactions.

With these parameters fixed, we predict from Eq. \ref{eqsupp}
an {\bf enhancement} factor
for $\J$ production of  
$1.2 < S_{\J} < 5.5$, where this range of values includes the
full range of initial parameters, i.e. initial temperatures between
300 and 500 MeV and charm quark rapidity ranges between 1 and 4.  
This is to be compared with predictions of models which
extrapolate existing suppression mechanisms to RHIC conditions, resulting
in typical suppression factors of 0.05 for central collisions \cite{Vogt}.
 Note that in addition to the qualitative change between suppression and
enhancement, the actual numerical difference should be easily detectable
by the RHIC experiments.  

One can also predict how this new effect will vary with the
centrality of the collision, which has been a key
feature of deconfinement signatures analyzed at
CERN SPS energies \cite{NA50}.  
To estimate the centrality dependence, we repeat
the calculation of the $\epsilon$ and $\beta$ parameters using
appropriate variation of initial conditions with impact parameter b.
From nuclear geometry and the total non-diffractive nucleon-nucleon
cross section at RHIC energies, one can estimate the total
number of participant nucleons $N_P(b)$ and the corresponding
density per unit transverse area $n_P(b,s)$ \cite{wounded}.
The former quantity has been shown to be directly
proportional to the total transverse energy produced in
a heavy ion collision \cite{ET}.  The latter quantity is used, along
with the Bjorken-model estimate of initial energy density
 \cite{Bj}, to provide an estimate of how the initial temperature
of the deconfined region varies with impact parameter.  We
also use the ratio of these quantities
to define an initial transverse area 
within which deconfinement is possible, thus completing the
initial conditions needed to calculate the $\J$ production
and suppression.  The average initial charm number
$\NZ$ varies with impact parameter in proportion to the
nuclear overlap integral $T_{AA}$(b).  The impact-parameter
dependence of the fraction $x$ is determined by the
average path length encountered by initial $\J$ as they
pass through the remaining nucleons, $L(b)$ \cite{gerschel}.
All of these b-dependent effects are normalized to the previous values 
used for calculations at $b = 0$.  

\begin{figure}[t]
\centerline{  \hspace*{-.cm}
\psfig{width=8.5cm,figure=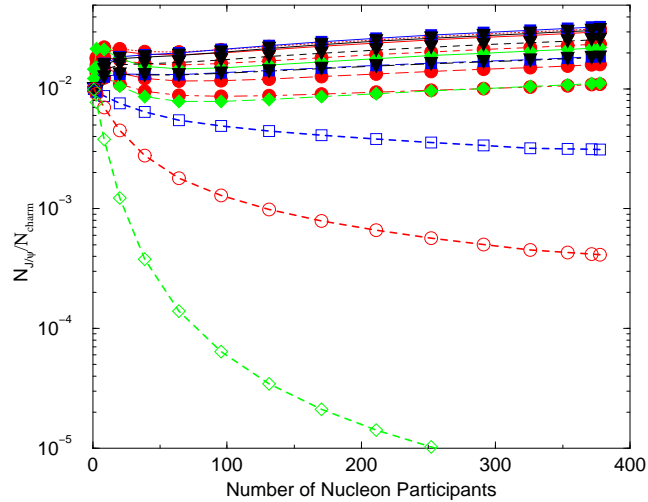}
}
\caption{ \small
 Ratio of final $\J$ to initial charm as a function of 
centrality.  The solid symbols are with the inclusion of the
new formation mechanism in the deconfined medium.  Initial
plasma temperature $T_0$ is coded by squares (300 MeV),
circles (400 MeV), and diamonds (500 MeV).  Different charm quark
momentum distributions are indicated by the solid lines (thermal) and
combinations of dashed and dotted lines ($\Delta$y = 1,2,3,4).  The
alternate screening model results are given by triangles, using
$T_s$ = 200, 240, and 280 MeV.  Curves with open symbols are the
same calculations with the new formation mechanism omitted.
\label{figjpsib}}
\end{figure}

It is revealing to express these
results in terms of the ratio of final $\J$ to initially-produced
charm pairs, both of which will be measurable at RHIC.  (This 
normalization automatically eliminates the trivial effects of 
increased collision energy
and phase space.)  
In Fig. \ref{figjpsib}, the solid symbols are the full results predicted 
 with the inclusion of 
our new production mechanism. We include full variation of these results
with initial temperature (squares, circles, and
diamonds are $T_0$ = 300, 400, 500 MeV, respectively), 
charm quark distribution (full lines are
thermal, combinations of dashed and dotted lines use $\Delta$y ranging
from one through 4), and 
also variation with screening temperature for the alternate scenario 
(triangles with $T_s$ = 200, 240, and 280 MeV).

The centrality
dependence is represented by the total
participant number $N_P$(b). The effect is somewhat obscured by the
log scale, but the ratio predictions typically increase  
about 50\% between
peripheral and central events.  Note that this increase is in addition
to the expected dependence of total charm production on centrality, so
that the quadratic nature of our new production mechanism is 
evident.

We also show for contrast the 
results without the new mechanism, when only dissociation by gluons is
included ($\lambda_F$ = 0 for curves with open symbols).  
These results have the opposite 
centrality dependence, and the absolute magnitudes are very much smaller.
It is evident that the
new mechanism dominates $\J$ production in a deconfined medium
at all but the largest impact parameters, and that this situation
survives uncertainties associated with variation in model parameters.  

For completeness, we list a few effects of variations in our
other parameters and assumptions which have relatively minor
impact on the results.

1.  The initial charm production at RHIC could be decreased due to
nuclear shadowing of the gluon structure functions.  Model estimates
\cite{gluonshadow}
indicate this effect could result in up to a 20\% reduction. 

2. The validity of the cross section used assumes strictly 
nonrelativistic bound states. 
Several alternative models for this cross section result in
substantially higher values.  When we arbitrarily 
increase the cross section
by a factor of two, or alternatively set the cross section to its
maximum value (1.5 mb) at all energies, 
we find an increase in the final
$\J$ population of about 15\%.  This occurs because the kinetics
always favors formation over dissociation, and a larger cross section
just allows the reactions to approach completion more easily within
the lifetime of the QGP.

3.  A nonzero transverse expansion will be expected at some level, 
which will reduce the lifetime of the QGP and reduce the 
efficiency of the new formation mechanism.  We have calculated
results for central collisions with variable transverse expansion, and
find a decrease in the parameter $\beta$ of about 15\% for each
increase of 0.2 in the transverse velocity.

4.  Model calculations of the approach to chemical equilibrium 
for light quarks and gluons indicate that the 
initial density of gluons in 
a QGP fall substantially below that for full phase space occupancy.
We have checked our model predictions in this scenario, using
a factor of two decrease in the gluon density at $\tau_0$.
This decreases the effectiveness of
the dissociation process, such that the final $\J$ production
is increased by about 35\%.  We also justify neglecting dissociation
via collisions with light quarks in this scenario, since the population
ratio of quarks to gluons is expected to be a small fraction.  This is
potentially important, since the inverse process is inhibited by
the required three-body initial state.

5.  The effect of a finite $\J$ formation time may also be considered.
The total effect is a competition between delayed dissociation ($\J$ cannot
be dissociated before formation) and a possible loss of states whose
formation started just before the hadronization point.  A conservative
upper limit would bound any decrease by the ratio of formation to
QGP lifetime, certainly in the 15\% range.

6.  Although the new formation mechanism is large compared with dissociation,
it is small on an absolute basis, with $\J$ yields only a few percent of
total charm.  These small values can be traced in part to the magnitude of
spatial charm density, which enters in the calculation of time-integrated
flux of charm quark pairs.  Our assumption of constant spatial density 
certainly underestimates the charm density, since it is likely 
somewhat peaked toward the center of the nuclear overlap region in each
collision.  A correspondingly smaller deconfined region is also to be
expected, but it will still contain virtually all initial charm and have
a similar time and longitudinal expansion profile.  
Thus a more realistic spatial
model should increase the formation yield beyond our simple estimates.

Overall, we predict that at high energies the $\J$ production
rate will provide an even better signal for deconfinement than
originally proposed.  Consideration of multiple heavy quark
production made possible by higher collision energy effectively
adds another dimension to the parameter space within which one
searches for patterns of quarkonium behavior in a 
deconfined medium. 

\begin{figure}[t]
\centerline{  \hspace*{-.cm}
\psfig{width=8.5cm,figure=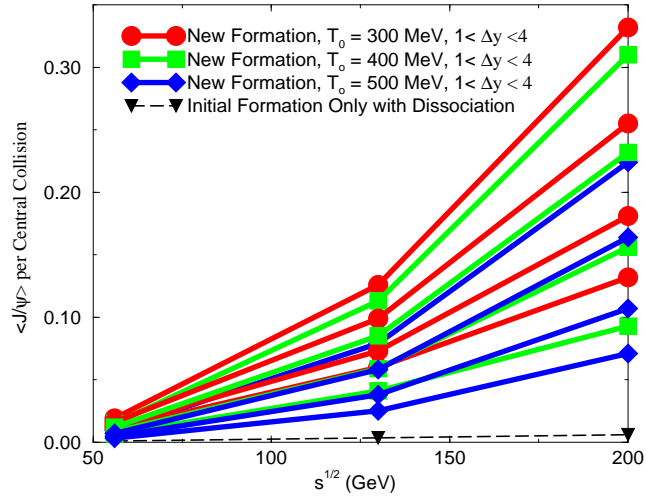}
}
\caption{ \small
 Predicted energy dependence of $\J$ at RHIC. 
\label{figjpsinewenergylinear}}
\end{figure}

The recent initial operation of RHIC at $\sqrt{s}$ = 56 and 130 GeV provides
an opportunity to test the predicted energy dependence of this new
mechanism.  We show in Fig. \ref{figjpsinewenergylinear} the expected 
energy variation
of the total $\J$ yield per central collision at RHIC.
The individual lines include full variation over the initial temperature
and charm quark momentum distributions.
The strong increase with energy comes from the quadratic dependence on 
initial charm production, coupled with the increase of the
charm production cross section with energy as calculated in 
pQCD \cite{hardprobes1}.  For comparison, we show the energy dependence
which results from just initial production, followed by dissociation alone.
If such a strong increase is observed at RHIC, 
it would signal the existence of a production mechanism
 nonlinear in initial charm.

Taken together, the enhanced magnitude  and centrality and energy
dependence 
predict signals which will be 
difficult to imitate with conventional hadronic
processes.  The extension of this scenario to LHC
energies will involve hundreds of initially-produced
charm quark pairs, and we expect the effects of this
new production mechanism to be striking.

This work was supported  by a grant from 
the U.S. Department of Energy, DE-FG03-95ER40937.


\end{document}